\documentclass[12pt]{article}
\pdfoutput=1
\usepackage{a4wide,graphicx,amsmath,amssymb}
\newcommand{\bear}{\begin{eqnarray}}
\newcommand{\eear}{\end{eqnarray}}
\usepackage[figuresright]{rotating}
\begin{document}

\title{\bf{Design of time delayed chaotic circuit with threshold controller}}

\author{K.~Srinivasan$^{1}$, I.~Raja Mohamed$^{2}$, K.~Murali$^3$, \\
M.~Lakshmanan$^{1,*}$ and Sudeshna Sinha$^4$  \\ \\
$^1$Centre for Nonlinear Dynamics, School of Physics, \\
Bharathidasan University, Tiruchirapalli - 620 024, India\\ \\
$^2$Department of Physics, B.S.Abdur Rahman University, Chennai-600 048, India\\ \\
$^3$Department of Physics, Anna University, Chennai-600 025, India \\ \\
$^4$Institute of Mathematical Sciences, C.I.T. Campus, Chennai-600 113, India and \\ Indian Institute of Science Education and Research, \\ 
Mohali Transit Campus, CHANDIGARH 160 019, India}
\date{\today}
\maketitle
Email : lakshman@cnld.bdu.ac.in \\

Running Title : Design of time delayed chaotic circuit with threshold controller \\


$^*$Corresponding author : M. Lakshmanan  
\newpage
\begin{abstract} 

A novel time delayed chaotic oscillator exhibiting mono- and double scroll complex chaotic attractors is designed.  This circuit consists of only a few operational amplifiers and diodes and employs a threshold controller for flexibility.  It efficiently implements a piecewise linear function.  The control of piecewise linear function facilitates controlling the shape of the attractors.  This is demonstrated by constructing the phase portraits of the attractors through numerical simulations and hardware experiments. Based on these studies, we find that this circuit can produce multi-scroll chaotic attractors by just introducing more number of threshold values.
 
\end{abstract}

\section{\label{sec:level1}Introduction} 

Nonlinear systems with time delayed feedback have attracted considerable attention due to their technological importance and their wide abundance in nature [see for example, Lu {\it{et al}}., 1998; Senthilkumar \& Lakshmanan, 2005; Senthilkumar {\it{et al}}., 2006].  Further, complex chaotic attractors have been obtained experimentally using simple electronic circuits in recent years [Namaj$\bar{u}$nas {\it{et al}}., 1995a; Tama$\check{s}$evi$\check{c}$ius {\it{et al}}., 2006; Wang \& Yang, 2006; Yalcin \& $\ddot{O}$zoguz, 2007; Wagemakers {\it{et al}}., 2008].  While the earlier studies on such delay systems mainly concentrated on constructing suitable nonlinear devices, in this letter we describe circuit implementation of nonlinear time delay systems which are easily controllable using a threshold mechanism so as to exhibit double scroll and even multiscroll chaotic attractors.  The simplest delay differential equation (DDE) describing a time delayed feedback oscillator can be written in the form
\begin{equation}
\frac{dx}{dt} = -ax(t) + bF[x(t-\tau)],
\label{gen_eq}
\end{equation}
where $a$ and $b$ are positive scalar parameters, $x(t)$ is a dynamical variable, $F(x)$ is a nonlinear activation function and $\tau$ is the time delay.  The well known   Mackey-Glass (MG) system [Mackey \& Glass, 1977; Namaj$\bar{u}$nas {\it{et al}}., 1995a] corresponds to the form
\begin{equation}
F(x)=F_1(x)=\frac{Ax}{1+x^m}.
\label{mg_eq}
\end{equation}
The parameters are commonly set at $A=2$ and $m=10$.  The MG model has been studied numerically by Farmer [1982], and also by Grassberger and Procaccia [1983], wherein they have obtained Lyapunov exponents, entropies and various types of dimensions.  It has also been studied experimentally using an electronic circuit to understand its dynamics [Namaj$\bar{u}$nas {\it{et al}}., 1995a, 1997] and for applications like controlling [Namaj$\bar{u}$nas {\it{et al}}., 1995b, 1997], synchronization [Tama$\check{s}$evi$\check{c}$ius {\it{et al}}., 1998; Sano {\it{et al}}., 2007; Kim {\it{et al}}., 2006] and secure communication [Pyragas, 1998].  Though the function $F_1(x)$ in (\ref{mg_eq}) exhibits an odd symmetry, the variable $x(t)$ in (\ref{gen_eq}) oscillates either in the positive ($x>0$) or in the negative ($x<0$) region depending on the initial conditions.  It does not switch between the two regions and the system exhibits only simple mono-scroll chaotic behaviour.  Other types of nonlinear functions $F(x)$ have also been  considered for Eq.~(\ref{gen_eq}).  These include piecewise constant rectangular-shaped function proposed by Haiden and Mackey and considered in [Losson {\it{et al}}., 1993; Schwarz \& M$\ddot{o}$gel, 1994], a piecewise linear tent-shaped function [Schwarz \& M$\ddot{o}$gel, 1994; Mykolaitis {\it{et al}}., 2003], a bell-shaped shifted Gaussian function [Voss, 2001] and a third-order polynomial function [Voss, 2002].  We note that all the above mentioned nonlinear functions exhibit only mono-scroll chaotic attractors.

Further, a five-segment function was introduced in [Lu \& He, 1996; Lu {\it{et al}}., 1998; Thangavel {\it{et al}}., 1998; Senthilkumar \& Lakshmanan, 2005]:
\begin{eqnarray}
F(x)=F_2(x)=
\left\{
\begin{array}{cc}
0, & \;\;\;\; \;\; x \le -\frac{4}{3} \\
-1.5Ax-2.0, & \;\;\;\; \;\; -\frac{4}{3}<x\le -0.8 \\
Ax, & \;\;\;\;\;  -0.8<x\le0.8 \\
-1.5Ax+2.0, & \;\;\;\; \;\; 0.8<x\le\frac{4}{3} \\ 
0 & \;\;\;\; \;\; x > \frac{4}{3}
\end{array} \right.
\label{than_eq}
\end{eqnarray}
where $A=1.0$, which also exhibits mono-scroll chaotic attractor.  The corresponding delay system has been studied in detail for bifurcations and chaos in [Senthilkumar \& Lakshmanan, 2005; Senthilkumar {\it{et al}}., 2006].  

Interestingly, with a slight alteration of (\ref{than_eq}), system (\ref{gen_eq}) was shown to possess a two-scroll chaotic attractor for $a = b = 1, \tau=8$, where the function $F(x)$ assumes a piecewise linear activation function of the form
\begin{eqnarray}
F(x)=F_3(x)=
\left\{
\begin{array}{cc}
B(x+1)-A & \;\;\;\; \;\;  x < -1, \\
Ax &\;\;\;\;\;  -1 \le x \le 1, \;\;\; (A>0, B<0) \\
B(x+1)+A &\;\;\;\; \;\;  x>1 
\end{array} \right.
\end{eqnarray}
where $A=2.15, B=-4.3$ [Tama$\check{s}$evi$\check{c}$ius {\it{et al}}., 2006].  Wang and Yang [2006] have shown the construction of multi-scroll chaotic attractor for the function $F(x)$ of the form
\begin{equation}
F(x)=F_4(x)=Ax+0.5(A-B)[(|x+m|-|x-m|)-(|x+n|-|x-n|)], \;\; (A>0, B<0)
\label{wang_eq}
\end{equation}
where $A=4.3, B=-5.8, m=1.1, n=3.3$.  Recently Yalcin \& $\ddot{O}$zoguz [2007] studied Eq.~(\ref{gen_eq}) experimentally by introducing delay in both the terms on the right hand side with the following form of nonlinear function $F(x)$, 
\begin{subequations} 
\begin{equation}
F(x)=F_5(x)=\sum_{i=1}^{M_x}g_{(-2i+1)/2}(x)+\sum_{i=1}^{N_x}g_{(2i-1)/2}(x),
\end{equation}
where
\begin{eqnarray}
g_{\theta}(\zeta)=
\left\{
\begin{array}{cc}
1, & \;\;\;\;    \zeta \geq \theta, \theta > 0, \\
0, & \;\;\;\;    \zeta < \theta, \theta > 0, \\
0, & \;\;\;\;    \zeta \geq \theta, \theta < 0, \\
-1, & \;\;\;\;    \zeta < \theta, \theta < 0.
\end{array} \right.
\end{eqnarray}
\label{yalcin_eq}
\end{subequations}
The authors found that the system exhibits $n$-scroll chaotic attractors for suitable values of $M_x$ and $N_x$.

In this letter, a piecewise linear function \emph{with a threshold controller} is introduced in the DDE instead of the commonly used $F(x)$ function in the literature as pointed above.  Depending on the delay parameter $\tau$ the novel system exhibits one-scroll and two-scroll chaotic attractors through the familiar period doubling route. We also discuss the hardware implementation of the system.  The experimental observations reveal the fact that this  circuit can also produce multi-scroll chaotic attractors by increasing the number of threshold voltage values. In particular, this method is effective and simple to implement since we only need to monitor a single state variable and reset it if it exceeds the threshold and so has potential engineering applications for various chaos-based information systems.

The activation function $F(x)$ in our study takes a symmetric piecewise linear form,
\begin{subequations} 
\begin{equation}
F(x)=F_6(x)=AF^* - Bx.
\end{equation}
Here
\begin{eqnarray}
F^*=
\left\{
\begin{array}{cc}
-x^* & \;\;\;\; \;\;  x < -x^*, \\
x &\;\;\;\;\;  -x^* \le x \le x^*, \\
x^*&\;\;\;\; \;\;  x>x^*, 
\end{array} \right.
\end{eqnarray}
\label{th_eq}
\end{subequations}
where $A$ and $B$ are positive parameters and $x^*$ is the controllable threshold value.  In our analysis, we chose $x^*=0.7, A=5.2, B=3.5$.  It may be noted that for $|x|>x^*$, the function $F_6(x)$ has a negative slope $-B$ and lies in all the four quadrants of the ($F-x$) plane which is shown in Fig.~1(a).  The figure reveals the piecewise linear nature of the function.  Experimental implementation of the function $F_6(x)$ is shown in Fig.~1(b) in the form of $V_{out}$ against $V_{in}$ characteristic by the nonlinear device unit $ND$ of Fig.~$2$.

\section{\label{cr1}Circuit realization}
The system described by Eq.~(\ref{gen_eq}) with the nonlinear function $F_6(x)$ can be constructed using analog electronic devices.  The circuit (Fig.~$2$) has a ring structure and comprises of a diode based nonlinear device unit (ND) with amplifying stages (U2, U3), a time delay unit (DELAY) with a buffer (U4) and an amplifying stage (U5), and a low pass first-order $R_0C_0$ filter.  The experimental time delay $T_D$ is given by
\begin{equation}
T_D=n\sqrt{LC}, \;\; n\ge 1
\label{exp_del_eq}
\end{equation}
where $n$ is the number of $LC$ filters.  The dimensionless delay parameter can be calculated by the relation
\begin{equation}
\tau=T_D/R_0C_0, \;\; n\ge 1.
\label{num_del_eq}
\end{equation}

The experimental circuit parameters are $R1 = 1~k\Omega, R2 = R3 = 10~k\Omega, R4 = 2~k\Omega, R5 = 3.0~k\Omega, R6 =10.4~k\Omega (\text{trimmer-pot}), R7 = 1~k\Omega, R8 = 5~k\Omega (\text{trimmer-pot}), R9 = R10 = 1~k\Omega, R11 = 10~k\Omega, R12 = 20~k\Omega (\text{trimmer-pot}), R_0 = 1.86~k\Omega, C_0 = 100~nF, L_i = 12~mH (i=1,2,...,11), C_i = 470~nF (i=1,2,...,10), n = 10$.  From (\ref{exp_del_eq}) and (\ref{num_del_eq}), we can see that $T_D = 0.751$ ms, $R_0C_0 = 0.19$ ms, and so $\tau = 3.952$.  In our circuit, $\mu$A741s are employed as operational amplifiers.  The constant voltage sources are $V1$, and $V2$, and the voltage supply for all active devices is $\pm12$ Volts.  The threshold value of the three segments involved in Eq.~(\ref{th_eq}) can be altered by adjusting the values of the voltages $V1$ and $V2$.

\section{\label{cr}Dynamics of the Time Delayed System with Threshold Controller}
\subsection{Period doubling route to chaos}
To start with, Eq.~(\ref{gen_eq}) has been integrated with the nonlinear function $F_6[x(t-\tau)]$ for the parametric values $a=1.0, b=1.2, A=5.2, B=3.5, x^*=0.7$, $\tau \in [1,3]$, where $\tau$ is employed as a control parameter.  The results are presented in Fig.~$3$ with the projection of phase trajectories.  For small values of the delay parameter $\tau$ the system exhibits period-T limit cycle.  Further increase of $\tau$ makes the system to undergo period-doubling bifurcations resulting in a chaotic attractor (Fig.~$3$).  For a range of $\tau$ values, $(1.3<\tau <1.98)$, mono-scroll attractors are formed similar to the MG and other delayed-feedback systems.  The phase trajectories stay on the mono-scroll isolated attractors up to $\tau=1.98$.  When $\tau$ is in the region, $1.98< \tau <6.5$, two-scroll chaotic attractors are formed and in this range some narrow periodic windows are also observed.  The dynamics of the circuit can also be studied experimentally through the associated analog circuit simulation (Fig.~$2$) and the results are illustrated in Fig.~$4$, where one division in horizontal and vertical axes correspond to $0.2~V$ and $2~V$, respectively.  From Figs.~$3$ and $4$, we observe that for the system (\ref{gen_eq}) and (\ref{th_eq}), the dynamical behaviour of the experimental circuit is similar to that obtained by numerical simulations.

It is to be mentioned that all the results shown in this paper are plotted after leaving out a very large number transients of the order $10^8$.  The details of bifurcations and chaos can be easily summarised by the one parameter bifurcation diagram in the $(\tau - x_{max})$ plane given in Fig.~$5(a)$. The period-doubling bifurcation sequence to chaos is observed for initial range of delay values (see Fig.~$5(b)$).  For instance, it is clear that for $1 < \tau < 1.15$ there is a limit-cycle attractor of period $T$.  At $\tau = 1.151$, a period-doubling bifurcation occurs and a period $2T$ limit cycle develops and is stable in the range $1.151 < \tau < 1.23$.  When the delay is increased further the period $2T$ limit cycle bifurcates to a period $4T$ attractor, and then to $8T$ and $16T$ period limit cycles.  After successive bifurcations, it forms a one-band chaotic attractor at $\tau = 1.26$.  On further increase in the delay value, the system exhibits double-band chaotic attractor.  From the bifurcation diagram, it may be noted that this interval of $\tau \in(3.1,5)$ is not fully occupied by the chaotic orbits alone.  Many fascinating changes in the dynamics take place at different critical values of $\tau$.  Particularly, the asymptotic motion consists of chaotic orbits interspersed by periodic orbits (windows) as may be seen from Fig.~$5$.  For $\tau > 5$ the system exhibits more complex hyperchaotic motion as in evident from the fact that more than one Lyapunov exponents are positive (see Fig.~$6$).  

\subsection{Lyapunov exponents and hyperchaotic regimes}
One of the interesting aspects of the dynamics associated with Eq.~(\ref{gen_eq}) and (\ref{th_eq}) is the existence of hyperchaos in a single first order scalar equation with time delay even for small values of $\tau$ while the other system parameters are fixed.  As the delay parameter is increased, for most parameter values the dimension increases and the attractor generally becomes more complicated, thereby contributing to the hyperchaotic nature of the system, which is confirmed by the increasing number of positive Lyapunov exponents.  The first seven maximal Lyapunov exponents, for parameter values $a=1.0, b=1.2, x^*=0.7, A=5.2, B=3.5$ and $\tau \in (1,10)$, are shown in Fig.~$6$, where it is evident that the number of positive Lyapunov exponents increases with time delay $\tau$.  One can see from Fig.~$6$, simple chaotic oscillations appear at about $\tau=1.0$ (single positive LE), while at $\tau = 1.35$ the second positive LE emerges, indicating hyperchaotic behaviour.  In the region $\tau \in [3.2,3.6]$ and $\tau \in [4.0,5.0]$ there exits several periodic windows with zero LE.  Also there are weakly hyperchaotic domains with relatively low values of LE.  Eventually for $\tau > 5.0$ the number of the positive LE starts to grow rapidly and becomes $7$ for $\tau >9.0$.

\subsection{One and Two parameter bifurcation diagrams}
A two parameter $(A-B)$ bifurcation diagram for the range $A \in [3.0, 8.0]$ and $B \in [1.1, 1.6]$ when $a=1.0, b=1.2, x^*=0.7$ and $\tau = 2.8$, is shown in Fig.~$6$, which clearly brings out the behaviour of the time delayed system with threshold controller for the combined system parameters $A$ and $B$ in a $800 \times 800$ grid (Fig.~$7$).  In this phase diagram, each colored region represents a particular type of steady-state behaviour: for example, red, period-$1$ attractor; green, period-$2$ attractor; blue, period-$3$ attractor; $4$-magenta; $5$-cyan; $6$-yellow; $7$-copper; $12$-gray; chaos-black; and further periodic regions-white.  Various dynamical phenomena are observed for different ranges of parameters $A$ and $B$.  In the range $A\in[3.0,5.4]$, we observe periodic orbits of periods $T, 2T, 3T, 6T, 12T, \ldots$ which leads to chaotic behaviour as the value of $B$ is increased.  Note that the period doubling route to chaos is observed from period $3T$.  On the other hand in the range $A\in[5.5,8.5]$, we observe regular period doubling phenomenon, starting from period $T$ to $2T,\,4T, \ldots$ leading to chaos as the value of B is increased.  In the chaotic region, many periodic windows are also observed.

Figure~$8(a)$ shows the dynamics of the system at $B=1.5$ for the range of $A \in [2.4, 9.0]$.  As the value of $A$ increases the system exhibits period $T, 3T, 6T, \ldots$ orbits and finally becomes chaotic (see Fig.~$8(b)$).  In a similar fashion the dynamics is demonstrated in Fig.~$9(a)$ for different values of $B$ at $A=5.2$.  As the value of $B$ is increased the period $T$ bifurcates to $2T$ and further increase in $B$ leads to oscillations of period $3T$.  Any further increase in $B$ leads to chaotic behaviour through novel period doubling route to chaos (see Fig.~$9(b)$).

\subsection{Amplification of chaotic attractors}
The system (\ref{gen_eq}) exhibits two-scroll chaotic attractor for $\tau=2.36$ with fixed values of other parameters as mentioned in the previous section.  In this section, the effect of the threshold controller is demonstrated for two values of $x^*$.  When $x^*=0.7$ the system (\ref{gen_eq}) exhibits a double band chaotic attractor.  For $x^*=1.1$ the system also exhibits a double band chaotic attractor with amplification of the oscillations.  The numerical and experimental results are shown in Fig.~$10$.  Note that the amplification is due to the fact that the threshold controller effectively controls the piecewise linear function which in turn alters the shape of the attractor and hence oscillations.  In short, threshold controlling mechanics facilitates the amplification of chaotic attractors.  Therefore, the controlling mechanism can help in altering the amplification of the signal depending upon the requirement.        

These details can be easily summarised by the one parameter bifurcation diagram in
the $(x^* - x_{max})$ plane given in Fig.~$11$.  From the figure, it is clear that an increase in the threshold value leads to an increase in the amplitude of the chaotic attractor.  

\section{\label{ml_cr}Multi-scroll Chaotic Attractors of the Time Delayed System with Threshold Controller}
In this section, we briefly demonstrate that the threshold controller can produce multi-scroll chaotic attractors also. The threshold control approach for creating multi-scroll attractors has  recently been studied in Jerk circuit [Lu {\it{et al}}., 2008].  Here, we show the effect of threshold controller in producing multi-scroll attractors in a time-delayed circuit for proper choice of the parameter values. Fixing $a=1.0, b=1.2, A=5.0, B=2.0$ and threshold values $x^*$ for the amplitude of the modulating square wave as $(0.7, 1.2)$,the system (\ref{gen_eq}) exhibits two single scroll chaotic attractors (Fig.~$12(Ia)$) when $\tau=1.4$ .  However when $\tau=2.2$  the system is found to exhibit two double scroll chaotic attractors (Fig.~$12(Ib)$).  

Correspondingly for the experimental investigation the threshold voltages $V1$ and $V2$ for the amplitude of the modulating square wave are chosen as $(0, 1.2~V)$.  The frequency of the square wave is fixed as  $100~Hz$.  The square wave used to modulate $V1$ switches between $0.0$ (which corresponds to $-0.7~V$ of diode break point voltage) and $-1.2~V$ and $V2$ switches between $0.0$ (which corresponds to $0.7~V$ of diode break point voltage) and $1.2~V$, respectively.  Figure~$12$ shows the plot of $(x(t-\tau), x(t))$, where the two single scroll chaotic attractor (Fig.~$12(IIa)$) and two double scroll chaotic attractor (Fig.~$12(IIb)$) are seen for the threshold voltages $V1$ and $V2$.   For these two values of threshold voltages, the chaotic attractor switches back and forth between the two breakdown voltages.  The jumping of the mono- and double scroll attractors between one chaotic state and another can be controlled by the frequency of the threshold modulated square wave.  Similarly multi-scroll chaotic attractors ($n>2$) can also be observed by choosing more number of threshold voltages.  That is, one needs to carefully clip the threshold function into multiple segments.

\section{Conclusion}

A novel time delayed chaotic oscillator exhibiting mono- and double scroll complex chaotic attractors is proposed.   This time delay system contains a threshold controlled piecewise nonlinearity.  By just adjusting the threshold values, the shape and size of the attractors can be varied.  Good qualitative agreement is obtained between the numerical simulation and the hardware experimental results.  The experiments reveal that this circuit can produce multi-scroll chaotic attractors by just adding more number of threshold voltages.  Due to its simplicity, this circuit can find  applications in communication and signal processing.

\section*{Acknowledgments} 
We acknowledge the help of Dr.D.V.Senthilkumar in the computing process.
This work is supported in part by the Department of Science and Technology (DST), Government of India – DST-IRHPA project, and DST Ramanna Fellowship of ML.

\newpage
\section*{References}
\begin{enumerate}

\item
Farmer, J. D. [1982]~``Chaotic attractor of an infinite dimensional dynamical system," {\it{Physica D}} {\bf 4}, 366-393.

\item
Grassberger, P. ~\&~ Procaccia, I [1983]~``Measuring the strangeness of strange attractors," {\it{Physica D}} {\bf 9}, 189-208.

\item
Kim, M.-Y., Sramek, C., Uchida, A. ~\&~ Roy, R. [2006] ~``Synchronization of unidirectionally coupled Mackey-Glass analog circuits with frequency bandwidth limitations," {\it{Phys. Rev. E}} {\bf 74}, 016211.

\item
Losson, J., Mackey, M. C. ~\&~ Longtin, A. [1993] ~``Solution multistability in first-order nonlinear differential delay equations," {\it{Chaos}} {\bf 3}, 167-176.

\item
Lu, H. ~\&~ He, Z. [1996] ~``Chaotic behavior in first-order autonomous continuous-time systems with delay," {\it{IEEE Trans. Circuits Syst. I: Fund. Th. Appl.}} {\bf 43}, 700-702.

\item
L$\ddot{u}$, J., Murali, K., Sudeshna Sinha, Henry Leung ~\&~ Aziz-Alaoui, M.A. [2008] ~``Generating multi-scroll chaotic attractors by thresholding," {\it{Phys. Lett. A}} {\bf 372}, 3234-3239.

\item
Lu, H., He, Y. ~\&~ He, Z. [1998] ~``A chaos-generator: analysis of complex dynamics of a cell equation in delayed cellular neural networks," {\it{IEEE Trans. Circuits Syst. I: Fund. Th. Appl.}} {\bf 45}, 178-181.

\item
Mackey, M. C. ~\&~ Glass, L. [1977]~``Oscillation and chaos in physiological control system," {\it{Science}} {\bf 197}, 287-289.

\item
Mykolaitis, G., Tama$\check{s}$evi$\check{c}$ius, A., $\check{C}$enys, A., Bumelien., S., Anagnostopoulos, A. N. ~\&~ Kalkan, N. [2003] ~``Very high and ultrahigh frequency hyperchaotic oscillators with delay line," {\it{Chaos Solit. Fract.}} {\bf 17}, 343-347.

\item
Namaj$\bar{u}$nas, A., Pyragas, K. ~\&~ Tama$\check{s}$evi$\check{c}$ius, A. [1995a] ~``An electronic analog of the Mackey-Glass system," {\it{Phys. Lett. A}} {\bf 201}, 42-46.

\item
Namaj$\bar{u}$nas, A., Pyragas, K. ~\&~ Tama$\check{s}$evi$\check{c}$ius, A. [1995b] ~``Stabilization of an unstable steady state in a Mackey-Glass system," {\it{Phys. Lett. A}} {\bf 204}, 255-262.

\item
Namaj$\bar{u}$nas, A., Pyragas, K. ~\&~ Tama$\check{s}$evi$\check{c}$ius, A. [1997] ~``Analog techniques for modelling and controlling the Mackey-Glass system," {\it{Int. J. Bifurcation and Chaos}} {\bf 7}, 957-962.

\item
Pyragas, K. [1998]~``Transmission of signals via synchronization of chaotic time-delay systems," {\it{Int. J. Bifurcation and Chaos}} {\bf 8}, 1839-1842.

\item
Sano, S., Uchida, A., Yoshimori, S ~\&~ Roy, R. [2007]~``Dual synchronization of chaos in Mackey-Glass electronic circuits with time-delayed feedback," {\it{Phys. Rev. E}} {\bf 75}, 016207(6).

\item
Schwarz, W. ~\&~ M$\ddot{o}$gel, A. [1994]~``Chaos generators with transmission line," {\it{Proc. 2nd Workshop on Nonlinear Dynamics of Electronics Systems, NDES'94}, Krakow}, 239-244.

\item
Senthilkumar, D. V. ~\&~ Lakshmanan, M. [2005] ~``Bifurcations and chaos in time delayed piecewise linear dynamical systems," {\it{Int. J. Bifurcation and Chaos}} {\bf 15}, 2895-2912.

\item
Senthilkumar, D. V., Lakshmanan, M. ~\&~ Kurths, J. [2006] ~``Phase synchronization in time-delay systems," {\it{Phys. Rev. E}} {\bf 74}, 035205(4).

\item
Tama$\check{s}$evi$\check{c}$ius, A., $\check{C}$enya, A., Namaj$\bar{u}$nas, A. ~\&~ Mykolaitis, G. [1998] ~``Synchronising hyperchaos in infinite-dimensional dynamical systems," {\it{Chaos Solit. Fract.}} {\bf 9}, 1403-1408.

\item
Tama$\check{s}$evi$\check{c}$ius, A., Mykolaitis, G. ~\&~ Bumeliene, S. [2006] ~``Delayed feedback chaotic oscillator with improved spectral characteristics," {\it{Electron. Lett.}} {\bf 42}, 736-737.

\item
Thangavel, P., Murali, K. ~\&~ Lakshmanan, M. [1998] ~``Bifurcation and controlling of chaotic delayed cellular neural networks," {\it{Int. J. Bifurcation and Chaos}} {\bf 8}, 24-81.

\item
Voss, H. U. [2001]~``A backward time shift filter for nonlinear delayed-feedback systems," {\it{Phys. Lett. A}} {\bf 279}, 207-214.

\item
Voss, H. U. [2002]~``Real-time anticipation of chaotic states of an electronic circuit," {\it{Int. J. Bifurcation and Chaos}} {\bf 12}, 1619-1625.

\item
Wagemakers, A., Buld$\acute{u}$, J. M. ~\&~ Sanju$\acute{a}$n, M. A. F. [2008] ~``Experimental demonstration of bidirectional chaotic communication by means of isochronal synchronization," {\it{Europhys. Lett.}} {\bf 81}, 40005.

\item
Wang, L. ~\&~ Yang, X. [2006] ~``Generation of multi-scroll delayed chaotic oscillator," {\it{Electron. Lett.}} {\bf 42}, 1439-1441.

\item
Yalcin, M. E., ~\&~ $\ddot{O}$zoguz, S. [2007] ~``n-scroll chaotic attractors from a first-order time-delay differential equation," {\it{Chaos}} {\bf 17}, 033112(8).

\end{enumerate}

\newpage
\section*{Caption of Figures}
\begin{enumerate}
\item[Figure 1:]
The nonlinear function $F_6(x)$. (a) Plot of the piecewise linear function $F_6(x)$ given by Eq.~(\ref{th_eq}). (b) Measured characteristic curve of the nonlinear device unit $ND$ from Fig.~2, $V_{out}$ against $V_{in}$.  Vertical scale $2~V/div$., horizontal scale $1~V/div.$

\item[Figure 2:]
Circuit diagram of the time delayed feedback oscillator with a nonlinear device unit $(ND)$, a time delay unit $(DELAY)$ and a lowpass first-order $R_0C_0$ filter.

\item[Figure 3:]
Period doubling bifurcations.  Phase portraits of Eqs.~(\ref{gen_eq}) with the function $F_6(x)$ given by Eq.~(\ref{th_eq}) for different values of the delay parameter: (a) $\tau=1.0$, Period-1 limit cycle; (b) $\tau=1.2$, Period-2 limit cycle; (c) $\tau=1.24$, Period-4 limit cycle; (d) $\tau=1.33$, One-scroll chaos; (e) $\tau=2.36$, Two-scroll chaos; (f) $\tau=2.8$, Two-scroll chaotic attractor.

\item[Figure 4:]
Phase portraits obtained from the experimental results of circuit of Fig.~$2$ for different delay parameter values corresponding to the numerical phase portraits of Fig.~$3$. $A=(R6/R4)=5.2$, $B=(R6/R5)=3.467$ and $\tau = T_D/R_0C_0$ : (a) $R_0=7500~\Omega$, Period-1 limit cycle; (b) $R_0=6250~\Omega$, Period-2 limit cycle; (c) $R_0=6050~\Omega$, Period-4 limit cycle; (d) $R_0=5640~\Omega$, One-scroll chaos; (e) $R_0=3180~\Omega$, Two-scroll chaos; (f) $R_0=2680~\Omega$, Two-scroll chaotic attractor.

\item[Figure 5:]
One parameter bifurcation diagram $(\tau - x_{max})$ for the parameter values $a=1.0, b=1.2, A=5.2, B=3.5, x^*=0.7$, (a) for $\tau \in (0.98,10)$ and (b) for the subset $\tau \in (1.0,1.6)$, after leaving out transients of order $1.0 \times 10^8$. 

\item[Figure 6:]
The first seven maximal Lyapunov exponents for parameter values $a=1.0, b=1.2, x^*=0.7, A=5.2, B=3.5$ and $\tau \in (1,10)$. 

\item[Figure 7:]
Two parameter $(A - B)$ bifurcation diagram in the range $A \in [3, 8.5]$ and $B \in [1.1, 1.6]$ when $a=1.0, b=1.2, x^*=0.7$ and $\tau = 2.8$, after leaving out transients of order $1.0 \times 10^8$.  The following color codes are used to represents various regions: red, period-$1$ attractor; green, period-$2$ attractor; blue, period-$3$ attractor; $4$-magenta; $5$-cyan; $6$-yellow; $7$-copper; $12$-gray; chaos-black; and further period regions-white. 

\item[Figure 8:]
One parameter bifurcation diagram $(A - x_{max})$ for the parameter values $a=1.0, b=1.2, \tau = 2.8, B=1.5, x^*=0.7$, (a) for $A \in (2.4,9)$ and (b) for the subset $A \in (2.4,4.5)$, after leaving out transients of order $1.0 \times 10^8$.

\item[Figure 9:]
One parameter bifurcation diagram $(B - x_{max})$ for the parameter values $a=1.0, b=1.2, A=5.2, \tau = 2.8, x^*=0.7$, (a) for $B \in (1.12,2)$ and (b) for the subset $B \in (1.12,1.4)$, after leaving out transients of order $1.0 \times 10^8$.

\item[Figure 10:]
Amplification of chaotic attractors. (I) Phase portraits (numerical) of two chaotic attractors of Eqs.~(\ref{gen_eq}) and (\ref{th_eq}) : (a) $x^* = 0.7$ and (b)   
$x^* = 1.1$. (II) Phase portraits (experimental) of two chaotic attractors from the circuit (Fig.~2), $x(t-\tau)$ against $x(t)$.  Vertical scale $2~V/div.$, horizontal scale $0.2~V/div.$: (a) $V1=V2=0.7~V$ and (b) $V1=V2=1.1~V$. $A=(R6/R4)=5.2$, $B=(R6/R5)=3.467$. 

\item[Figure 11:]
One parameter bifurcation diagram $(x^* - x_{max})$ for the parameter values $a=1.0, b=1.2, A=5.2, B=3.5, \tau = 2.8$ and $x^* \in (0.45,20)$. 

\item[Figure 12:] Multi-scroll chaotic attractors with threshold modulation of square wave amplitudes. (I) Phase portraits (numerical) of chaotic attractors of Eqs.~(\ref{gen_eq}) and (\ref{th_eq}) : 
(a) $\tau = 1.4$, two single scroll and (b) $\tau = 2.2$, two double scroll attractors.  (II) Phase portraits (experimental) of chaotic attractors from the circuit (Fig.~2), $x(t-\tau)$ against $x(t)$.  Vertical scale $2~V/div.$, horizontal scale $0.5~V/div.$: (a) $R_0 = 5364~\Omega$, two single scroll and (b) $R_0 = 3413~\Omega$, two double scroll attractors. $A=(R6/R4)=5.0$, $B=(R6/R5)=2.01$ and 
$\tau = T_D/R_0C_0$.

\end{enumerate}

\newpage
%

\begin{figure}
\centering
\includegraphics[width=0.8\columnwidth]{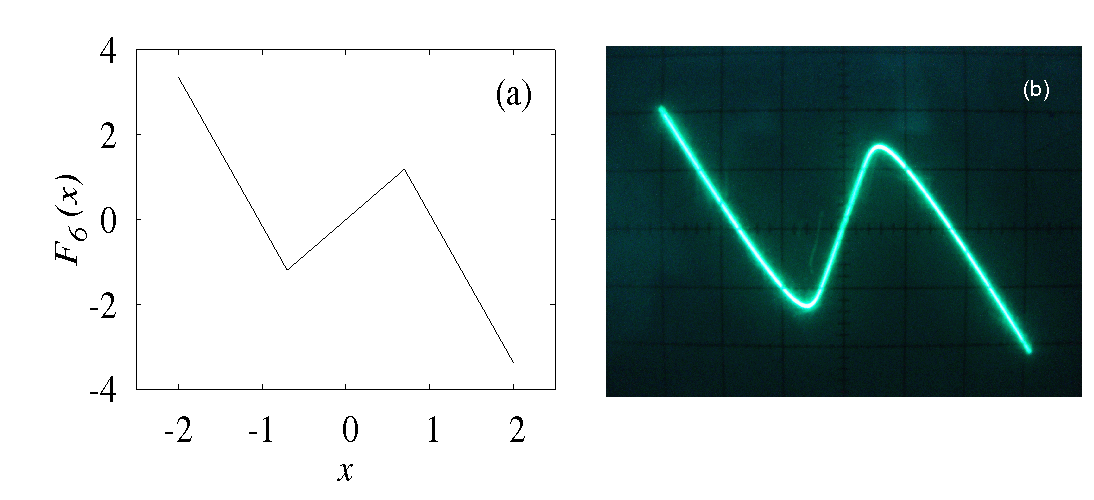}
\caption{\label{fig1} }
\end{figure}

\begin{figure}
\centering
\includegraphics[width=0.9\columnwidth]{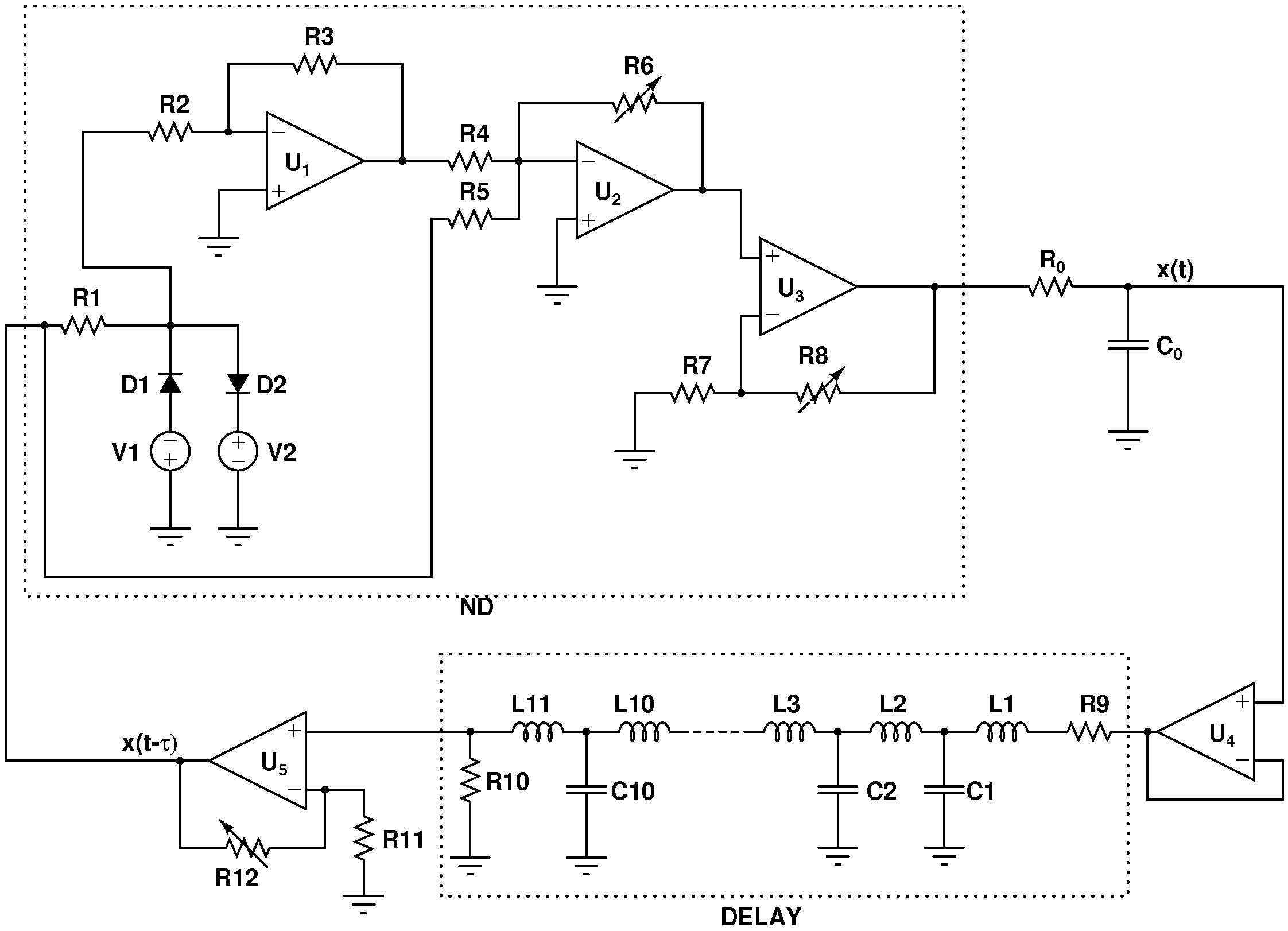}
\caption{\label{fig2} }
\end{figure}

\begin{figure}
\centering
\includegraphics[width=0.7\columnwidth]{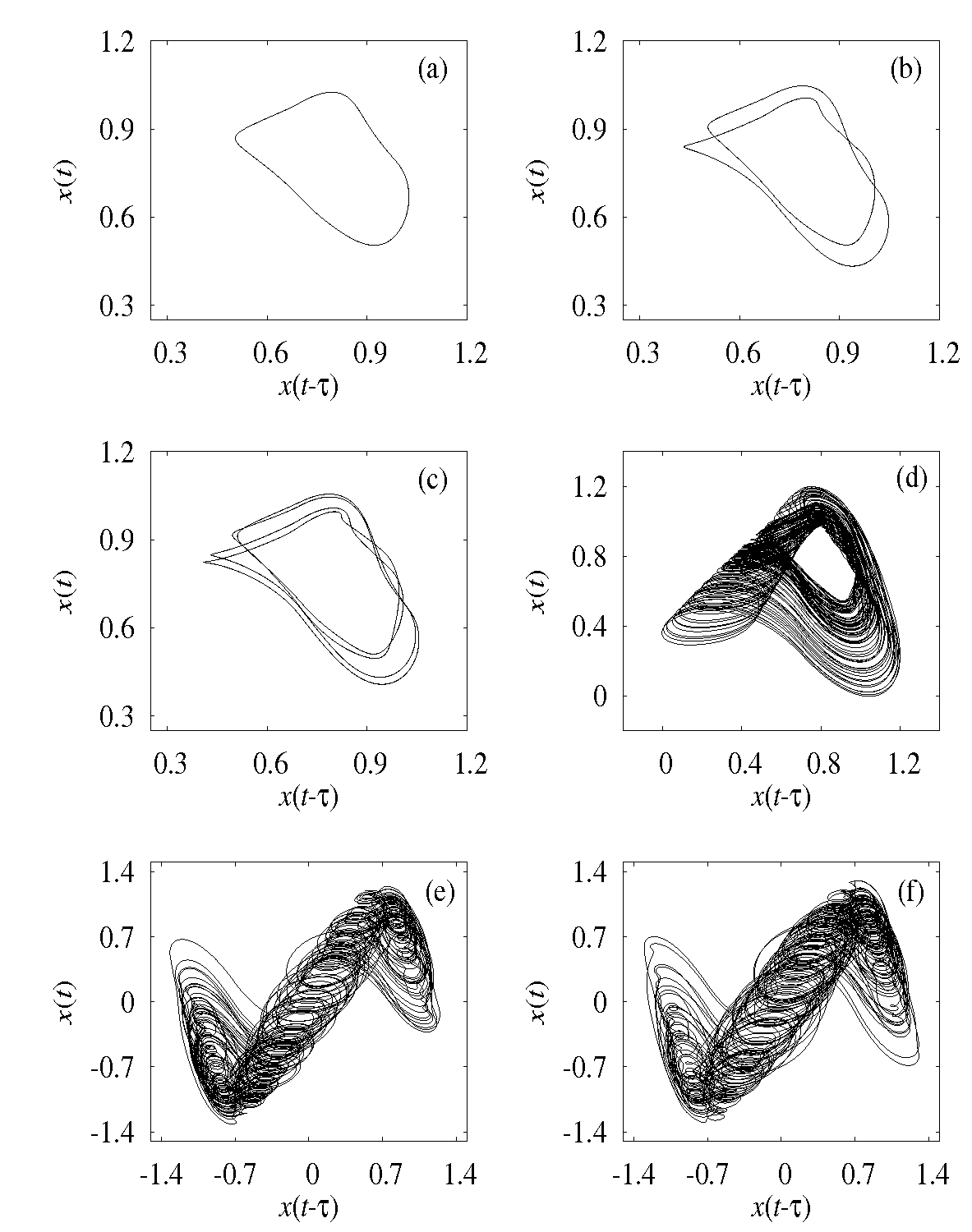}
\caption{\label{fig3} }
\end{figure}
\begin{figure}
\centering
\includegraphics[width=0.7\columnwidth]{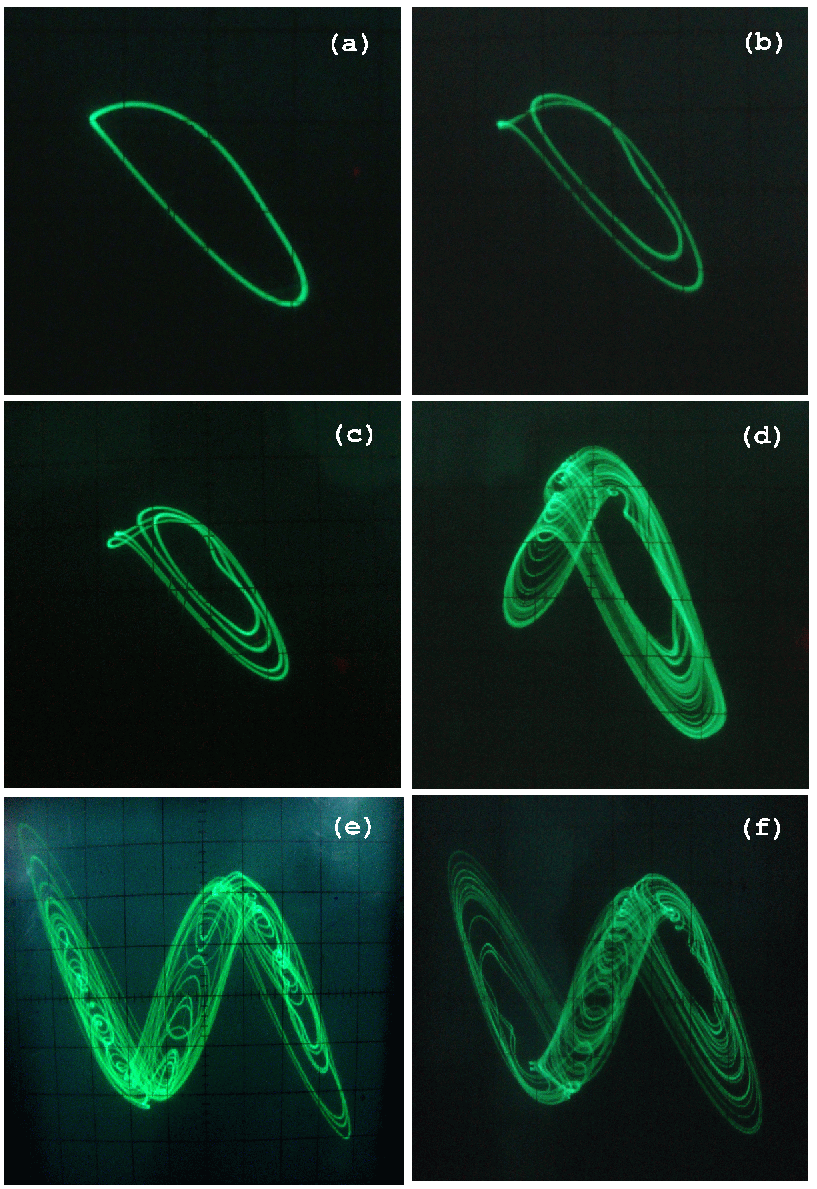}
\caption{\label{fig4} }
\end{figure}
\begin{figure}
\centering
\includegraphics[width=0.9\columnwidth]{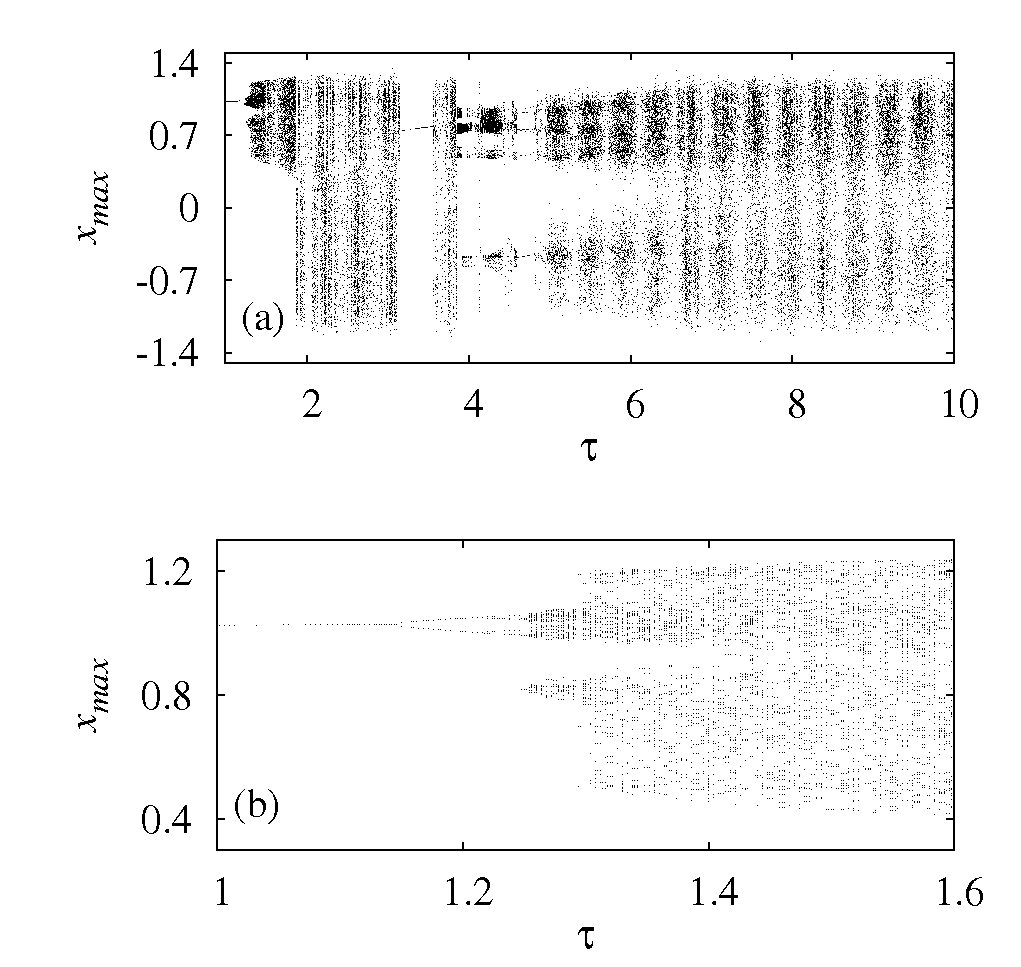}
\caption{\label{fig5} }
\end{figure}
\begin{figure}
\centering
\includegraphics[width=0.9\columnwidth]{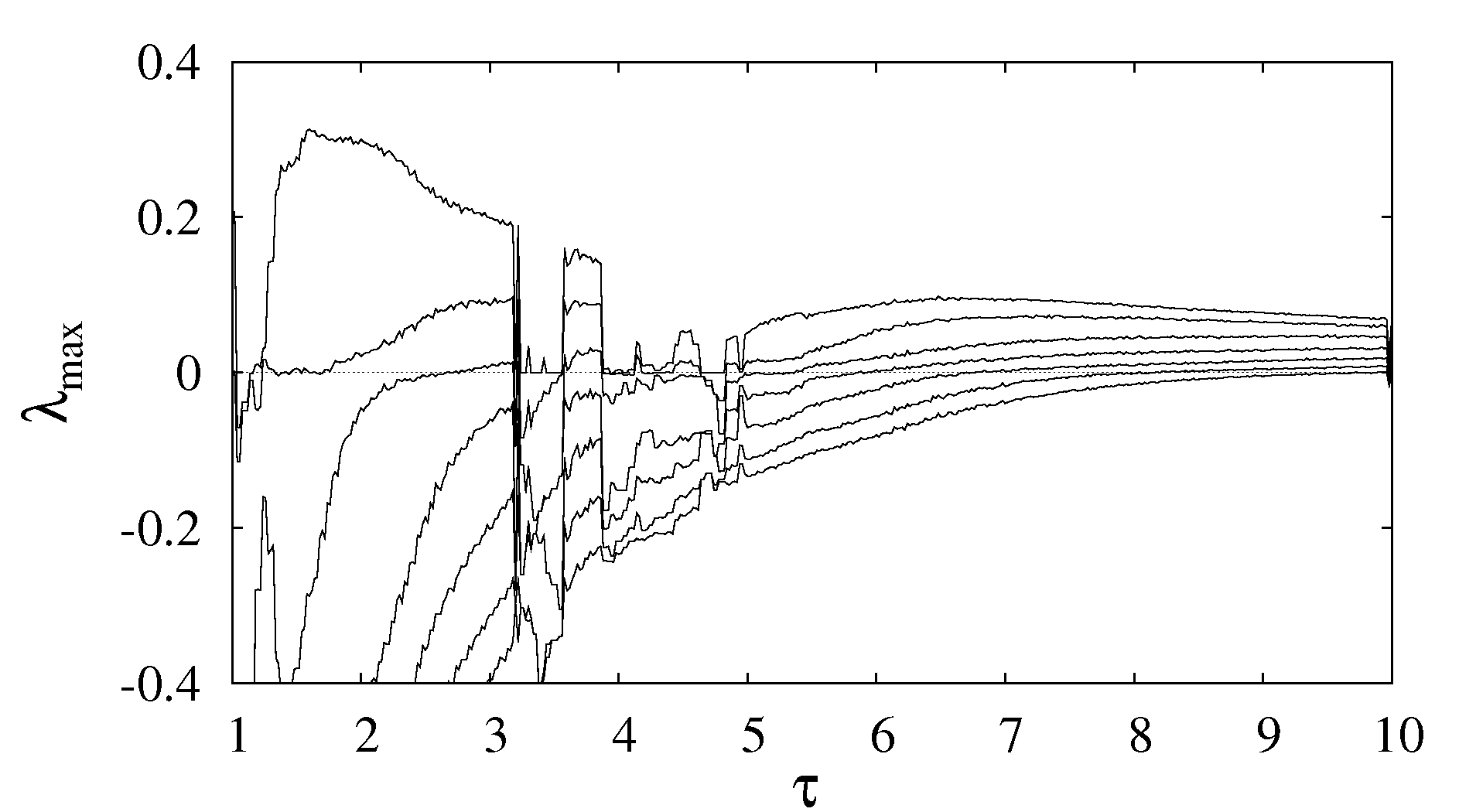}
\caption{\label{fig9} }
\end{figure}
\begin{figure}
\centering
\includegraphics[width=0.9\columnwidth]{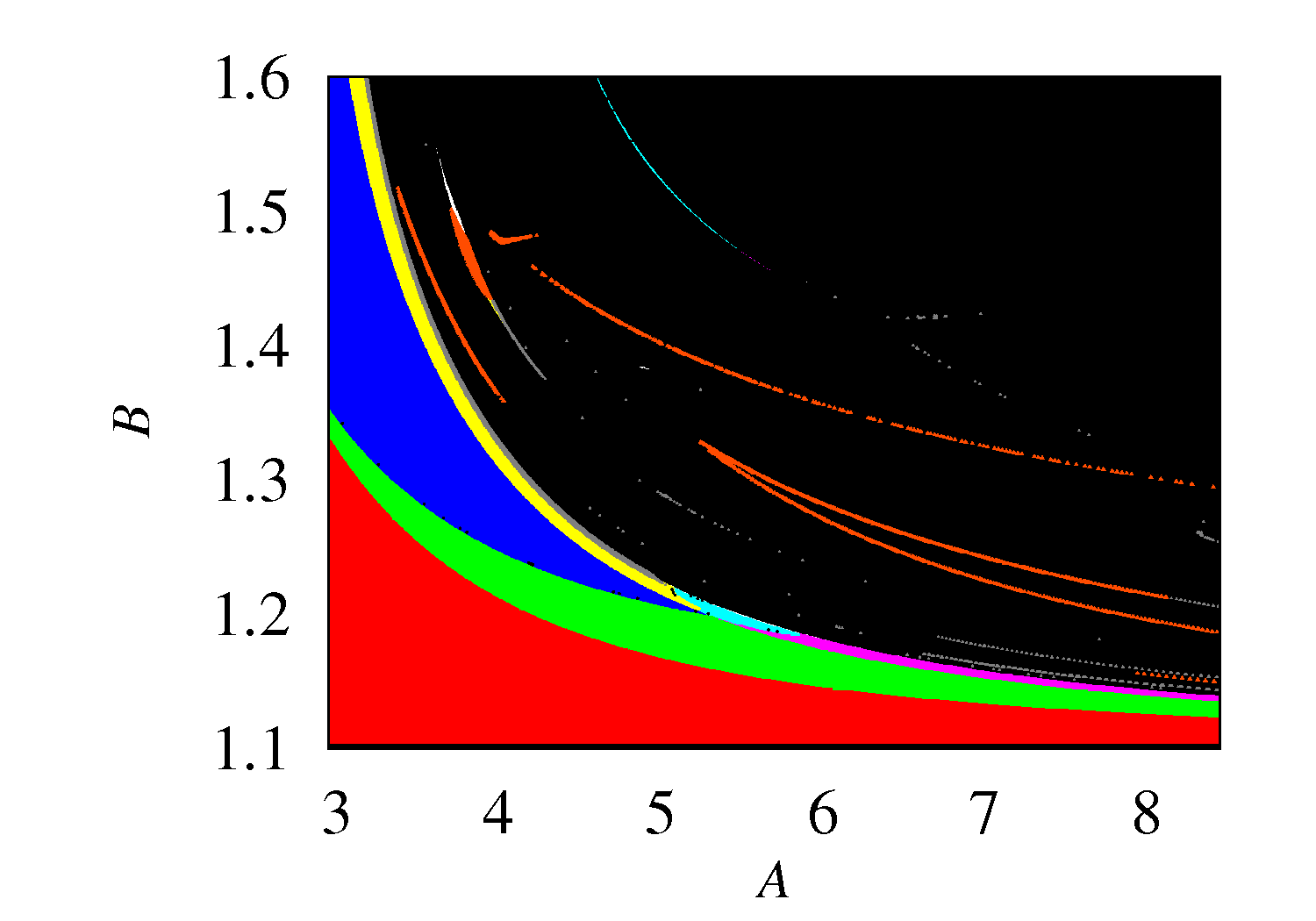}
\caption{\label{fig6} }
\end{figure}
\begin{figure}
\centering
\includegraphics[width=0.9\columnwidth]{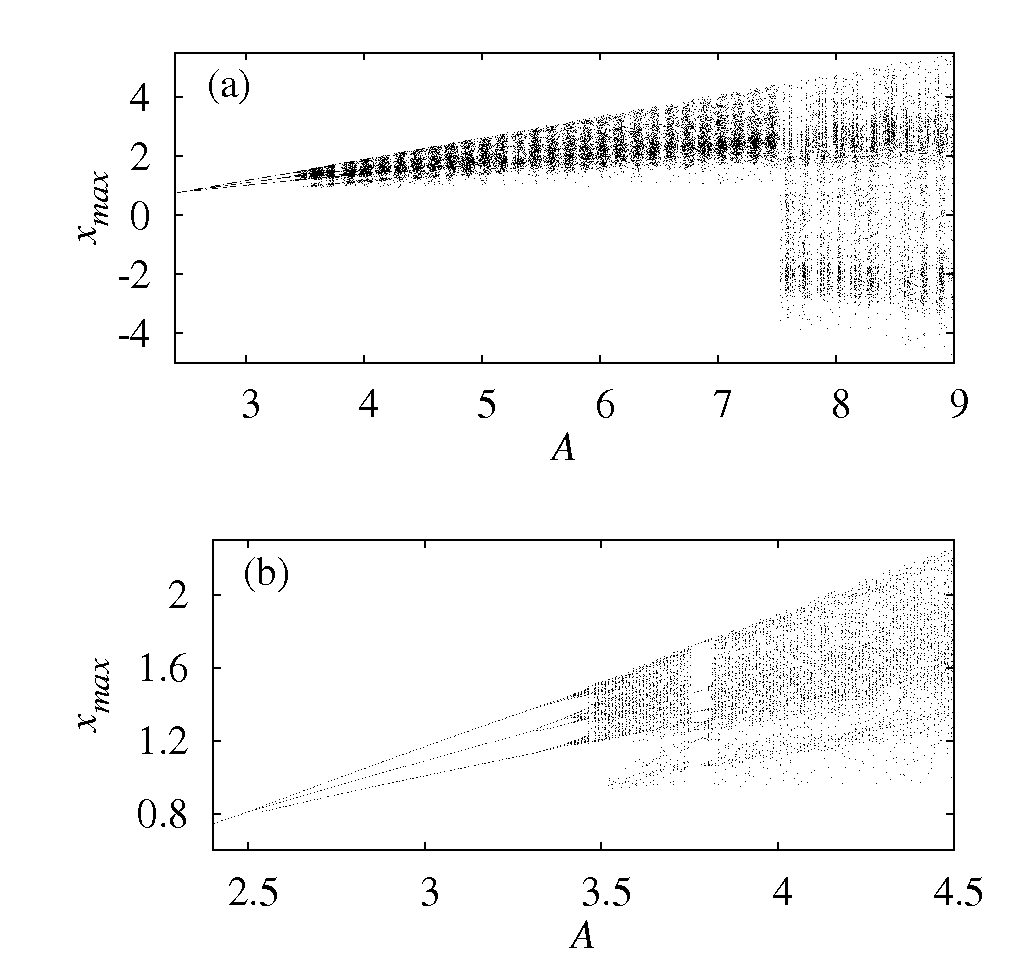}
\caption{\label{fig7} }
\end{figure}
\begin{figure}
\centering
\includegraphics[width=0.9\columnwidth]{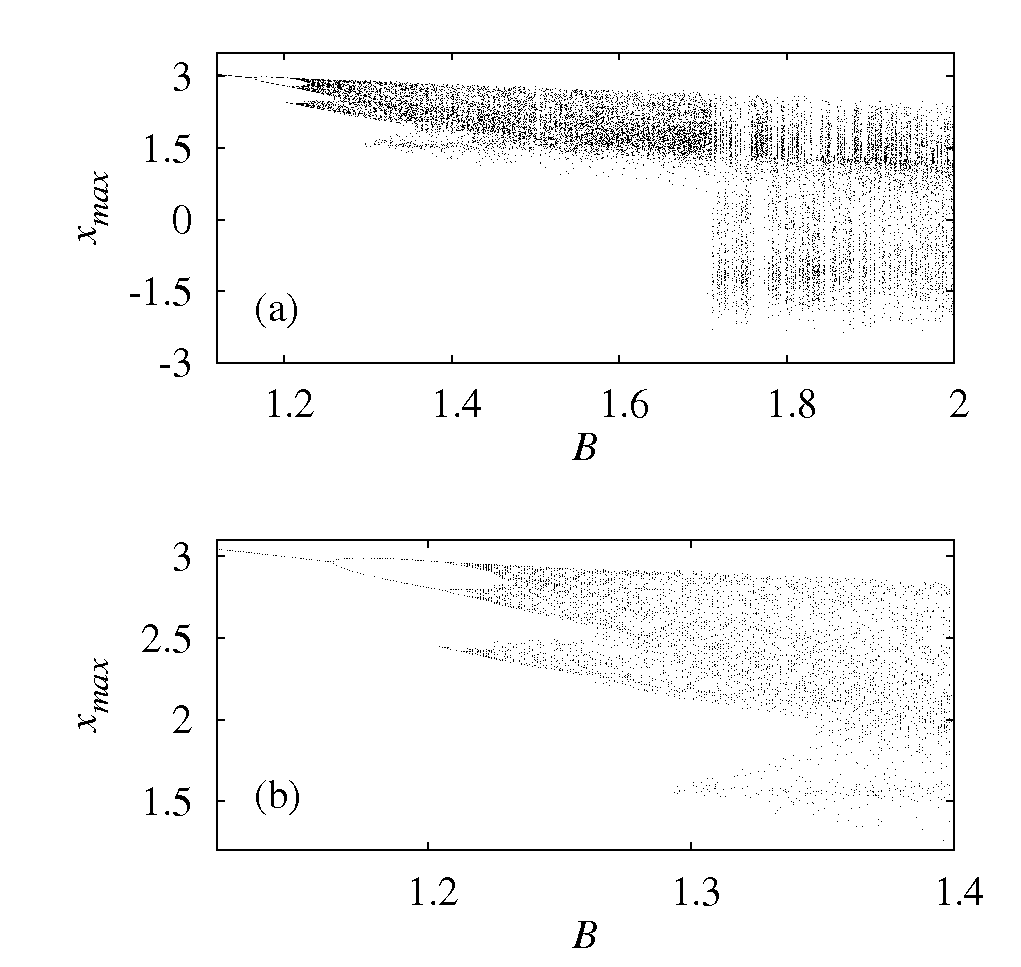}
\caption{\label{fig8} }
\end{figure}
\begin{figure}
\centering
\includegraphics[width=1.0\columnwidth]{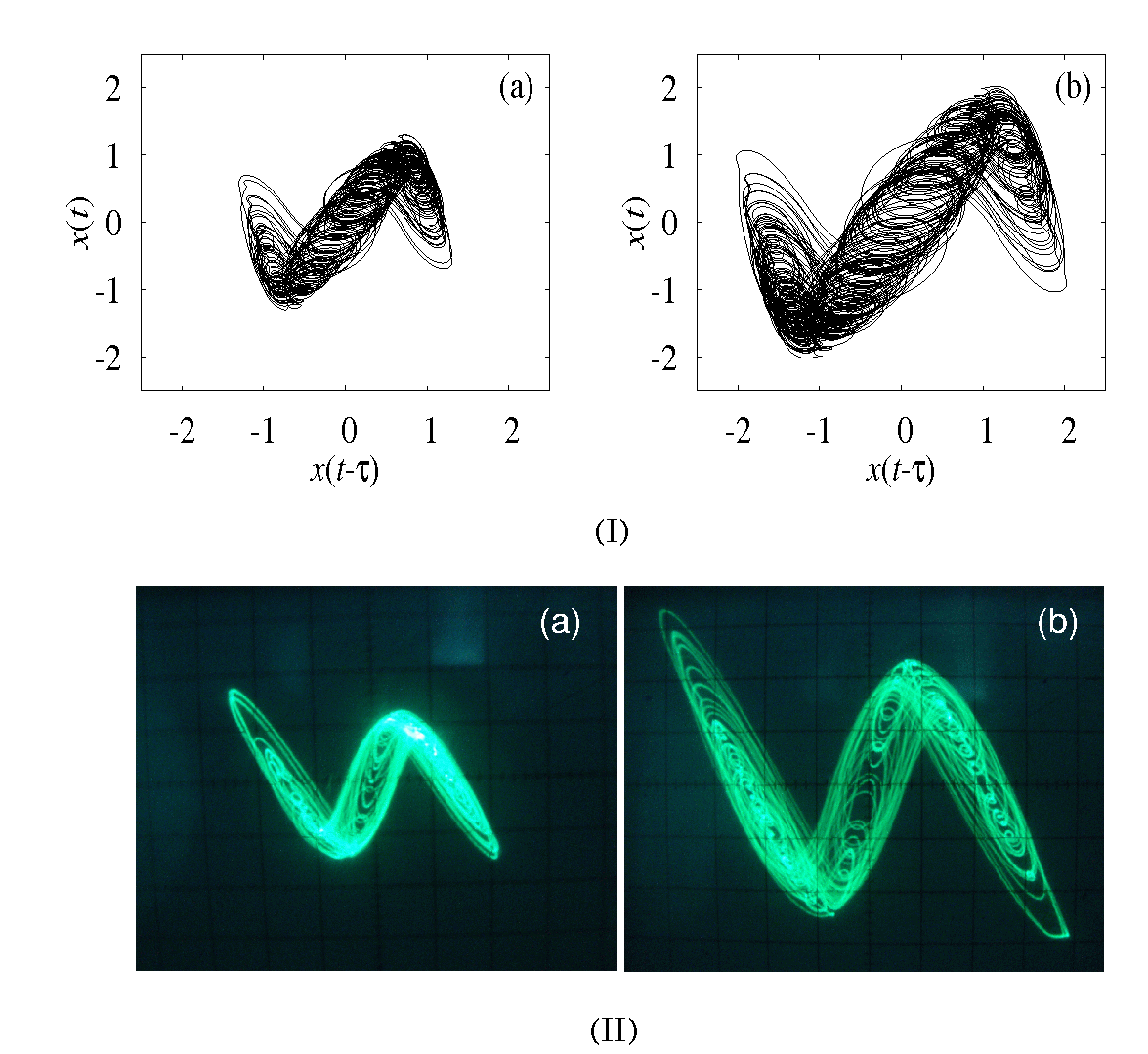}
\caption{\label{fig10} }
\end{figure}
\begin{figure}
\centering
\includegraphics[width=1.0\columnwidth]{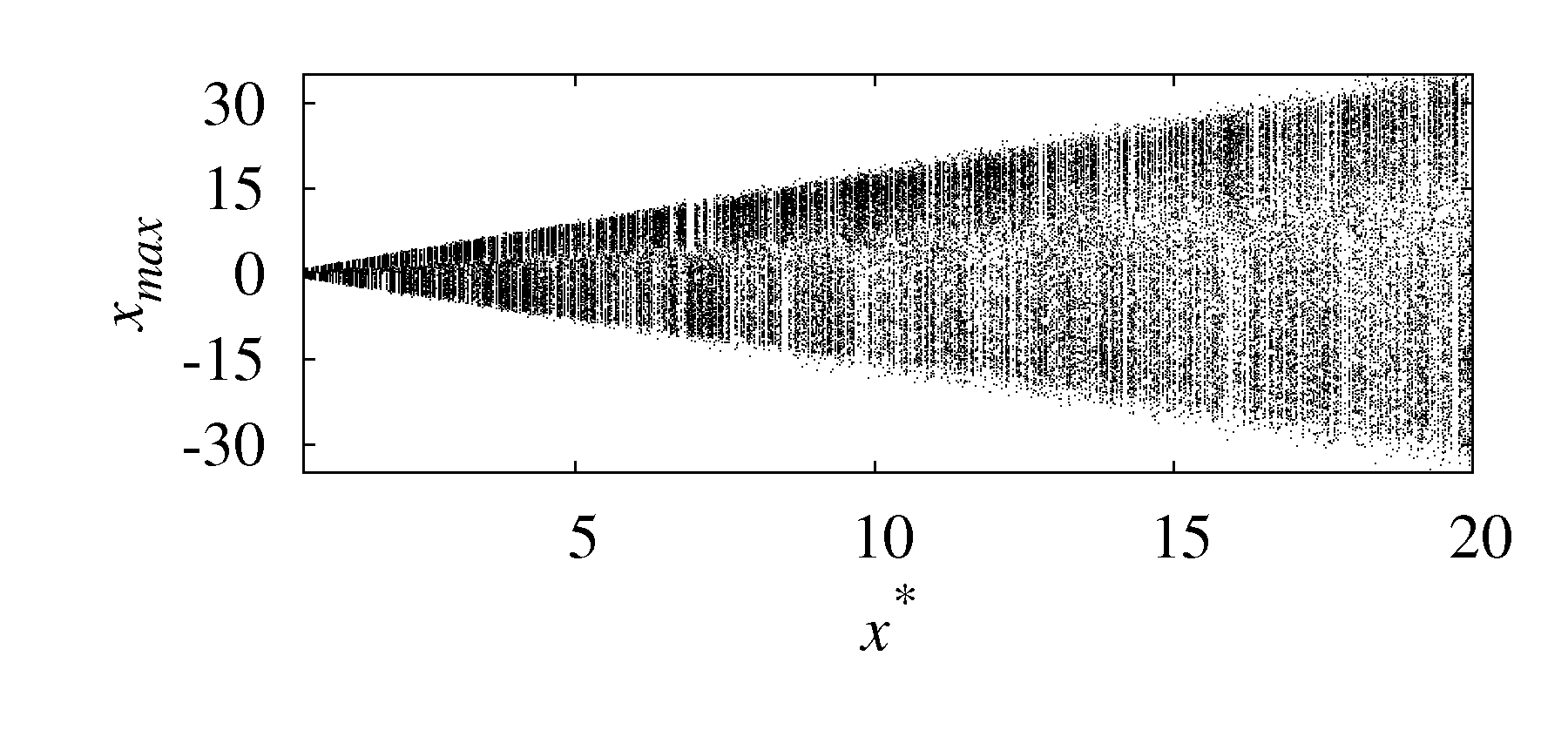}
\caption{\label{fig11} }
\end{figure}
\begin{figure}
\centering
\includegraphics[width=0.8\columnwidth]{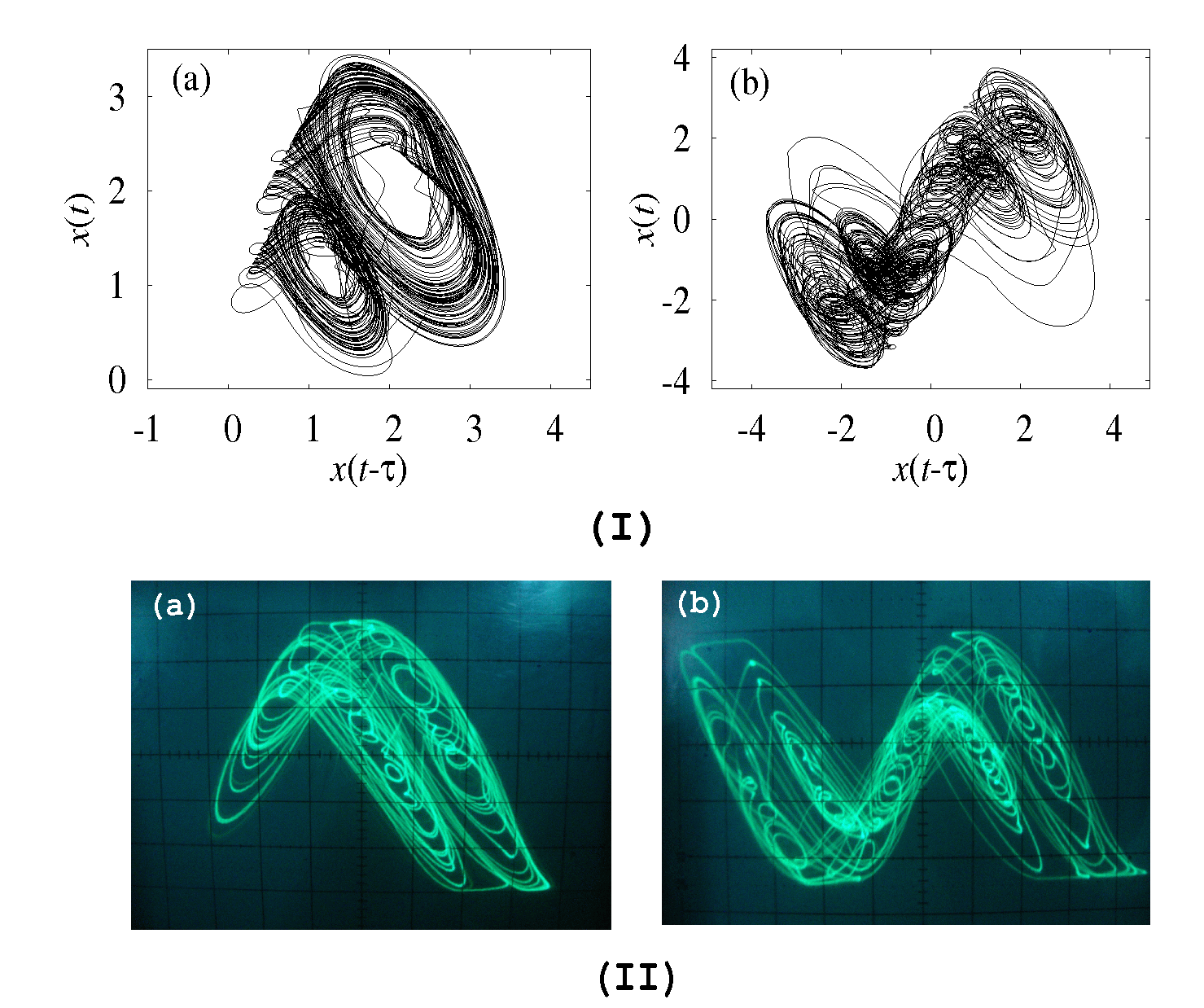}
\caption{\label{fig12} }
\end{figure}
%
%
\end{document}